\documentclass[preprint,aps,amsmath,showpacs,showkeys,nofootinbib, 
superscriptaddress]{revtex4} 
\usepackage[dvips]{graphicx} 
 
\begin{document} 

\title{\textbf{$\Omega d$ bound state}}
\author{H.~Garcilazo} 
\email{humberto@esfm.ipn.mx} 
\affiliation{Escuela Superior de F\' \i sica y Matem\'aticas, \\ 
Instituto Polit\'ecnico Nacional, Edificio 9, 
07738 M\'exico D.F., Mexico} 

\author{A.~Valcarce} 
\email{valcarce@usal.es} 
\affiliation{Departamento de F\'\i sica Fundamental,\\ 
Universidad de Salamanca, E-37008 Salamanca, Spain}

\date{\today} 

\begin{abstract} 
The lattice QCD analyses of the HAL QCD Collaboration
predicts a strongly attractive potential
in the $\Omega N$ $^5S_2$ channel which supports a bound state.
In this paper we show that this $\Omega N$ channel together
with the $NN$ $^3S_1$ channel
give rise to a $\Omega d$ bound state in
the state with maximal spin $(I,J^P)=(0,5/2^+)$ with a
binding energy of $\sim$ 17 MeV.
\end{abstract}



\maketitle 

\newpage 

The $\Omega N$ interaction is expected to lack a repulsive core 
since the quark flavors of the nucleon are different
from those of the omega so that the Pauli exclusion
principle can not act~\cite{Eti14}. The very recent lattice QCD
results for baryon-baryon interactions obtained at nearly physical 
quark masses by the HAL QCD Collaboration~\cite{DOI18} show that
the $\Omega N$ interaction in the $^5S_2$ channel is strongly
attractive such that it supports a bound state as indicated by
the effective-range expansion $k \, {\rm cot}\delta(k)/m_{\pi}$ 
at low energies. There are already estimations of observables 
in heavy-ion collisions from the experiments at RHIC and LHC 
to show experimentally the strong $p\Omega$ interaction~\cite{Mor16}.

In a very recent work~\cite{SEK18} Sekihara, Kamiya and Hyodo
have developed a model to study the origin of the attraction
in the $\Omega N$ $^5S_2$ interaction. 
They constructed an equivalent
local potential for the $\Omega N$ $^5S_2$ interaction
which reproduces the $\Omega N$ $^5S_2$ scattering amplitude
of Ref.~\cite{DOI18}. 
The long range part of the 
potential is based on $\eta$ meson exchange and correlated two mesons in
the scalar-isoscalar channel, denoted by "$\sigma$" in the
literature. The short range part is represented by a contact
interaction and they also added the box diagrams with intermediate
$\Lambda\Xi$, $\Sigma\Xi$, and $\Lambda\Xi(1530)$ channels where
the first two are inelastic ones so that they generate an imaginary
part for the potential.
Our purpose in this work is to make use
of that interaction together with the Malfliet-Tjon
model of Ref.~\cite{FRI90} for the $^3S_1$ $NN$ potential
to study the $\Omega d$ system in the maximal spin channel,
$(I,J^P)=(0,5/2^+)$.

It is worth to emphasize that $\Omega N$ system in the $^5S_2$ 
channel cannot couple to the lower channels $\Lambda\Xi$ or $\Sigma\Xi$ where these 
last ones are in a $S$ wave state, so that the width of the $\Omega N$
bound state is expected to be very small. 
Similarly, by choosing
the $\Omega d$ in the maximal spin channel $(I,J^P)=(0,5/2^+)$
the coupling to the lower channels $\Lambda\Xi N$ and $\Sigma\Xi N$
does not involve the $\Lambda\Xi$ and $\Sigma\Xi$ $S$ waves, so that the
width of a $\Omega d$ bound state is also expected to be small.

In order to solve the $\Omega NN$ three-body problem we use the 
method of Ref.~\cite{GAR16} where the two-body amplitudes are expanded in
terms of Legendre polynomials. Taking into account that two of
the particles are identical and
assuming that particle 1 is the $\Omega$ and 
the identical particles 2 and 3
are the two nucleons, there are only two independent interactions;
$V_1(r)$ which is the $NN$ interaction in the $^3S_1$ channel and
$V_2(r)$ which is the $\Omega N$ interaction in the $^5S_2$ channel.
Thus, the Faddeev equations for the bound state problem of the
$\Omega d$ system in the $(I,J^P)=(0,5/2^+)$ channel take the 
simple form~\cite{GAR16},
\begin{equation}
T_2^n(q_2)=\sum_m \int_0^\infty dq_3 K^{nm}(q_2,q_3;E)T_2^m(q_3) \, ,
\label{eq1}
\end{equation}
where
\begin{equation}
K^{nm}(q_2,q_3;E) =A^{nm}_{23}(q_2,q_3;E)
+\sum_l\int_0^\infty dq_1 A_{31}^{nl}(q_2,q_1;E)
A_{13}^{lm}(q_1,q_3;E) \, ,
\label{eq2}
\end{equation}
and
\begin{equation}
A^{nm}_{ij}(q_i,q_j;E) =
\sum_r \tau_i^{nr}(E-q_i^2/2\nu_i)\frac{q_j^2}{2}
\int_{-1}^1 d\,{\rm cos}\theta 
\frac{P_r(x'_i)P_m(x_j)}{E-p_j^2/2\eta_j-q_j^2/2\nu_j} \, ,
\label{eq3}
\end{equation}
\begin{equation}
\tau^{nr}_i(e)=\frac{2n+1}{2}\frac{2m+1}{2}
\int_{-1}^1 dx_i \int_{-1}^1 dx'_i\; 
P_n(x_i)t_i(x_i,x'_i;e)P_r(x'_i) \, ,
\label{eq4}
\end{equation}
\begin{equation}
x_i=\frac{p_i-b}{p_i+b} \, .
\label{eq5}
\end{equation}
$p_i$ and $q_i$ are the magnitude of the Jacobi relative momenta,
\begin{equation}
\vec p_i=\frac{m_k\vec k_j-m_j\vec k_k}{m_j+m_k},
\label{eq55}
\end{equation}
\begin{equation}
\vec q_i=\frac{m_i(\vec k_j+\vec k_k)-(m_j+m_k)\vec k_i}{m_i+m_j+m_k} \, ;
\label{eq56}
\end{equation}
while $\eta_i$ and $\nu_i$ are the corresponding reduced masses,
\begin{equation}
\eta_i=\frac{m_jm_k}{m_j+m_k},
\label{eq57}
\end{equation}
\begin{equation}
\nu_i=\frac{m_i(m_j+m_k)}{m_i+m_j+m_k}.
\label{eq58}
\end{equation}
$t_i(x_i,x'_i;e)$ corresponds to the off-shell two-body $t-$matrix
$t_i(p_i,p'_i;e)$ through the transformation~(\ref{eq5}),
with $b$ a scale parameter on which the solution does not depend.
The off-shell two-body $t-$matrices are obtained by solving
the Lippmann-Schwinger equation,
\begin{equation}
t_i(p_i,p_i^\prime;e)= V_i(p_i,p_i^\prime)+ \int_0^\infty 
{p_i^{\prime\prime}}^2dp_i^{\prime\prime}
V_i(p_i,p_i^{\prime\prime})
\frac{1}{e-{p_i^{\prime\prime}}^2/2\eta_i+i\epsilon} 
t_i(p_i^{\prime\prime},p_i^\prime;e) \, .
\label{eq6}
\end{equation}

In the case of the $NN$ subsystem the two-body potential
in configuration space is given by~\cite{FRI90},
\begin{equation}
V_1(r)=\sum_{n=1}^2C_n \frac{\exp(-\mu_n r)}{r} \, ,
\label{eq7}
\end{equation}
with $C_1=-626.885$ MeV fm, $C_2=1438.72$ MeV fm, $\mu_1=1.55$ fm$^{-1}$, 
and $\mu_2=3.11$ fm$^{-1}$. In momentum space we have,
\begin{equation}
V_1(p_1,p_1^\prime)=\frac{1}{2\pi p_1 p_1^\prime}\sum_{n=1}^2 C_n \,
ln  \left[\frac{(p_1+p_1^\prime)^2+\mu_n^2}
{(p_1-p_1^\prime)^2+\mu_n^2} \right] \, .
\label{eq8}
\end{equation}
In the case of the $\Omega N$ subsystem the two-body potential
in configuration space is given by~\cite{SEK18},
\begin{equation}
V_2(r)=\frac{\pi^3}{2}\sum_{n=1}^9 C_n
\left(\frac{\Lambda^2}{\Lambda^2-\mu_n^2}\right)^2
\left[\frac{\exp(-\mu_n r)}{r}-\frac{\exp(-\Lambda r)}{r}
-\frac{\Lambda^2-\mu_n^2}{2\Lambda}\exp(-\Lambda r)\right] \, ,
\label{eq9}
\end{equation}
with $C_n$ given by the real part of the last column 
in Table V of Ref.~\cite{SEK18}, $\Lambda=1$ GeV,
and $\mu_n=n\times 100$ MeV. In momentum space we have,
\begin{eqnarray}
V_2(p_2,p_2^\prime) & = & \frac{\pi^2}{4 p_2 p_2^\prime}\sum_{n=1}^9 C_n
\left(\frac{\Lambda^2}{\Lambda^2-\mu_n^2}\right)^2 
\left\{ ln \left[\frac{(p_2+p_2^\prime)^2+\mu_n^2}
{(p_2-p_2^\prime)^2+\mu_n^2} \right]     
-ln \left[\frac{(p_2+p_2^\prime)^2+\Lambda^2}
{(p_2-p_2^\prime)^2+\Lambda^2}\right] \right. \nonumber \\
& & + \left. \frac{\Lambda^2-\mu_n^2}{\Lambda^2+(p_2+p_2^\prime)^2}
-\frac{\Lambda^2-\mu_n^2}{\Lambda^2+(p_2-p_2^\prime)^2} \right\}.
\label{eq10}
\end{eqnarray}
With all these ingredients, we obtained a $\Omega d$ binding energy of 16.34 MeV
measured with respect to the $\Omega NN$ threshold. For comparison, the 
$\Omega N$ binding energy in this model is only 0.3 MeV \cite{SEK18}.
 
We also obtained the wave function of the $\Omega NN$ state and used it to
calculate the imaginary part of the eigenvalue using first-order perturbation
theory as,
\begin{equation}
\Delta E=\frac{\langle \Psi\mid\delta V\mid\Psi\rangle}
{\langle \Psi\mid\Psi\rangle}= -i\, 2.05\,\,{\rm MeV} \, .
\label{eq11}
\end{equation}
Using the same procedure we calculated the Coulomb energy as 0.85 MeV so that
the final value of the energy measured with respect to the
$\Omega NN$ threshold is:
\begin{equation}
E=-17.19
 -i\, 2.05\,\,{\rm MeV}.
\label{eq12}
\end{equation}

The recent developments of interactions in the strange sector based on lattice QCD
together with our findings for the $\Omega d$ system 
are an indication that may serve as a motivation 
for experimental searches. As discussed in Ref.~\cite{Mor16} there exist already observables
in heavy-ion collisions that may unveil the existence of these states. 
We hope our theoretical studies could help to design experiments where 
these lattice QCD based predictions could be tested. 

\begin{acknowledgments} 
This work has been partially funded by COFAA-IPN (M\'exico), 
by Ministerio de Econom\'\i a, Industria y Competitividad 
and EU FEDER under Contract No. FPA2016-77177-C2-2-P
and by Junta de Castilla y Le\'on under Contract No. SA041U16.
\end{acknowledgments}

\end{document}